\documentclass[letterpaper,twocolumn,superscriptaddress,floatfix]{revtex4}
\usepackage{inputenc}
\usepackage{bm}
\usepackage{multirow,amssymb,amsbsy,amsmath}
\usepackage{graphicx}
\usepackage{float}
\usepackage{verbatim}
\makeatletter
\usepackage{pifont}
\usepackage{upgreek}
\usepackage{color}
\usepackage{indentfirst}
\usepackage{soul}

\newcommand{\Rmnum}[1]{\expandafter\@slowromancap\romannumeral #1@}
\usepackage{caption}
\makeatother
\makeatletter 
\renewcommand{\section}{\@startsection{section}{1}{0mm}
	{-\baselineskip}{0.5\baselineskip}{\bf\leftline}}
\makeatother

\begin{document}
	\title{Maskless Generation of Single Silicon Vacancy Arrays in Silicon Carbide by a Focused He$^+$ Ion Beam}
	
	\affiliation{CAS Key Laboratory of Quantum Information, University of Science and Technology of China, Hefei, Anhui 230026, China}
	\affiliation{CAS Center For Excellence in Quantum Information and Quantum Physics,
		University of Science and Technology of China, Hefei, Anhui 230026, China}
	\affiliation{Hefei National Laboratory, University of Science and Technology of China, Hefei, Anhui 230088, China}
	\affiliation{Center for Micro and Nanoscale Research and Fabrication, University of Science and Technology of China, Hefei, Anhui 230026, People's Republic of China}
	\affiliation{State Key Laboratory of Functional Materials for Informatics, Shanghai Institute of Microystem and Information Technology, Chinese Academy of Science, Shanghai 20050 China}
	
	\author{Zhen-Xuan He}
	\altaffiliation{Equal contribution}
	\affiliation{CAS Key Laboratory of Quantum Information, University of Science and Technology of China, Hefei, Anhui 230026, China}
	\affiliation{CAS Center For Excellence in Quantum Information and Quantum Physics,
		University of Science and Technology of China, Hefei, Anhui 230026, China}
	
	\author{Qiang Li}
	\altaffiliation{Equal contribution}
	\affiliation{CAS Key Laboratory of Quantum Information, University of Science and Technology of China, Hefei, Anhui 230026, China}
	\affiliation{CAS Center For Excellence in Quantum Information and Quantum Physics,
		University of Science and Technology of China, Hefei, Anhui 230026, China}
	\affiliation{State Key Laboratory of Functional Materials for Informatics, Shanghai Institute of Microystem and Information Technology, Chinese Academy of Science, Shanghai 20050 China}
	
	\author{Xiao-Lei Wen}
	\affiliation{Center for Micro and Nanoscale Research and Fabrication, University of Science and Technology of China, Hefei, Anhui 230026, People's Republic of China}
	
	\author{Ji-Yang Zhou}
	\affiliation{CAS Key Laboratory of Quantum Information, University of Science and Technology of China, Hefei, Anhui 230026, China}
	\affiliation{CAS Center For Excellence in Quantum Information and Quantum Physics,
		University of Science and Technology of China, Hefei, Anhui 230026, China}
	
	\author{Wu-Xi Lin}
	\affiliation{CAS Key Laboratory of Quantum Information, University of Science and Technology of China, Hefei, Anhui 230026, China}
	\affiliation{CAS Center For Excellence in Quantum Information and Quantum Physics,
		University of Science and Technology of China, Hefei, Anhui 230026, China}
	
	\author{Zhi-He Hao}
	\affiliation{CAS Key Laboratory of Quantum Information, University of Science and Technology of China, Hefei, Anhui 230026, China}
	\affiliation{CAS Center For Excellence in Quantum Information and Quantum Physics,
		University of Science and Technology of China, Hefei, Anhui 230026, China}
	
	\author{Jin-Shi Xu}
	\altaffiliation{Email: jsxu@ustc.edu.cn}
	\affiliation{CAS Key Laboratory of Quantum Information, University of Science and Technology of China, Hefei, Anhui 230026, China}
	\affiliation{CAS Center For Excellence in Quantum Information and Quantum Physics,
		University of Science and Technology of China, Hefei, Anhui 230026, China}
	\affiliation{Hefei National Laboratory, University of Science and Technology of China, Hefei, Anhui 230088, China}

	\author{Chuan-Feng Li}
	\altaffiliation{Email: cfli@ustc.edu.cn}
	\affiliation{CAS Key Laboratory of Quantum Information, University of Science and Technology of China, Hefei, Anhui 230026, China}
	\affiliation{CAS Center For Excellence in Quantum Information and Quantum Physics,
		University of Science and Technology of China, Hefei, Anhui 230026, China}
	\affiliation{Hefei National Laboratory, University of Science and Technology of China, Hefei, Anhui 230088, China}
	\author{Guang-Can Guo}
	\affiliation{CAS Key Laboratory of Quantum Information, University of Science and Technology of China, Hefei, Anhui 230026, China}
	\affiliation{CAS Center For Excellence in Quantum Information and Quantum Physics,
		University of Science and Technology of China, Hefei, Anhui 230026, China}
	\affiliation{Hefei National Laboratory, University of Science and Technology of China, Hefei, Anhui 230088, China}
	
	\begin{abstract}
		Precise generation of spin defects in solid-state systems is essential for nano-structure fluorescence enhancement. We investigated a method for creating single silicon vacancy defect arrays in silicon carbide using a helium-ion microscope. Maskless and targeted generation can be realized by precisely controlling the focused He$^+$ ion beam with an implantation uncertainty of 60 nm. The generated silicon vacancies were identified by measuring the optically detected magnetic resonance spectrum and room temperature photoluminescence spectrum. We systematically studied the effects of implantation ion dose on the generated silicon vacancies. After optimization, a conversion yield of $\sim$ 6.95 \% and a generation rate for a single silicon vacancy of $\sim$ 35 \% were realized. This work paves the way for the integration and engineering of color centers to photonic structures and the application of quantum sensing based on spin defects in silicon carbide.

	\end{abstract}
	
	\maketitle
	\date{\today}
	\section*{Introduction}
	
	As a wide-band semiconductor material, silicon carbide (SiC) is known for its well-established inch-scale production, mature controlled-doping, high-quality nanofabrication, and CMOS-friendly process. Recent years, spin defects in SiC, including the negatively-charged silicon vacancy, neutral divacancy and negatively-charged nitrogen-vacancy, have become promising platform for quantum information processing \cite{koehl2011room,falk2013polytype,christle2015isolated,widmann2015coherent,nagy2018quantum,fuchs2015engineering,nagy2019high,yang2014electron,carter2015spin,son2020developing,christle2017isolated,morioka2020spin,Babin_2021,soykal2016silicon,wang2020coherent,mu2020coherent}. Among them, silicon vacancy (V$_{Si}$) has attracted great interest due to the long spin coherent times at room temperature \cite{widmann2015coherent,simin2017locking}, versatile applications in quantum sensing \cite{kraus2014magnetic,simin2015high,niethammer2016vector}, sharp zero-phonon lines in resonant excitation, \cite{nagy2018quantum,nagy2019high,banks2019resonant} and also the prospective applications in constructing spin-to-photon interface \cite{morioka2020spin,Babin_2021,soykal2016silicon}. Except the intrinsic defects, there are different approaches to generate silicon vacancies in SiC. Electron irradiation \cite{widmann2015coherent,nagy2018quantum} and neutron irradiation \cite{fuchs2015engineering} have been used to generate V$_{Si}$ with excellent optical and spin properties. However, the distribution of V$_{Si}$ defects is random, making it difficult to integrate color centers into microstructures like waveguides, nanobeams, and photonics crystals \cite{Babin_2021,radulaski2017scalable,bracher2017selective}. By employing PMMA mask and ions implantation, V$_{Si}$ can be generated with a precisely controlled position.\cite{Babin_2021,wang2017efficient,wang2019demand}. Nevertheless, the lithography process of the mask causes inconvenience, and the thickness of the mask limits the implantation depth. Methods have been developed to realized maskless creation of color centers, like laser writing \cite{chen2019laser}, proton beam lithography \cite{kraus2017three,ohshima2018creation} and focused ion beam (FIB) \cite{wang2017scalable,pavunny2021arrays}. Among these, FIB has become an advanced technology in on-demand defects engineering due to the precisely-controlled position of generated color centers, revealing great potential in the wafer-scale nanofabrication process of single-photon emitters\cite{hollenbach2022wafer}. Moreover, the helium-ion microscope (HIM), which possesses spatial controlled FIB technology and also a higher resolution (0.25 nm) than SEM, has been used for precise-imaging nanofabrication and generating defects in semiconductor materials like nitrogen-vacancy ensembles in diamond \cite{huang2013diamond,mccloskey2014helium}. 
	
	In this work, we utilized the focused ion beam of HIM to generate V$_{Si}$ defects ensembles and V$_{Si}$ defects arrays in 4H-SiC with high precision. The generated V$_{Si}$ defects are identified through the optically detected magnetic spectrum and room temperature photoluminescence spectrum. By optimizing the dose of the focused ion beam,  we realized a conversion yield of $\sim$ 6.95 \% and a generation rate for single silicon vacancy of $\sim$ 35 \%. Our method of on-demand and maskless generation of V$_{Si}$ defects would be used for nanoscale integration of spin defects for quantum photonics and quantum sensing applications.
	
	\begin{figure*}[ht]
		\centering
		\includegraphics[scale = 0.8]{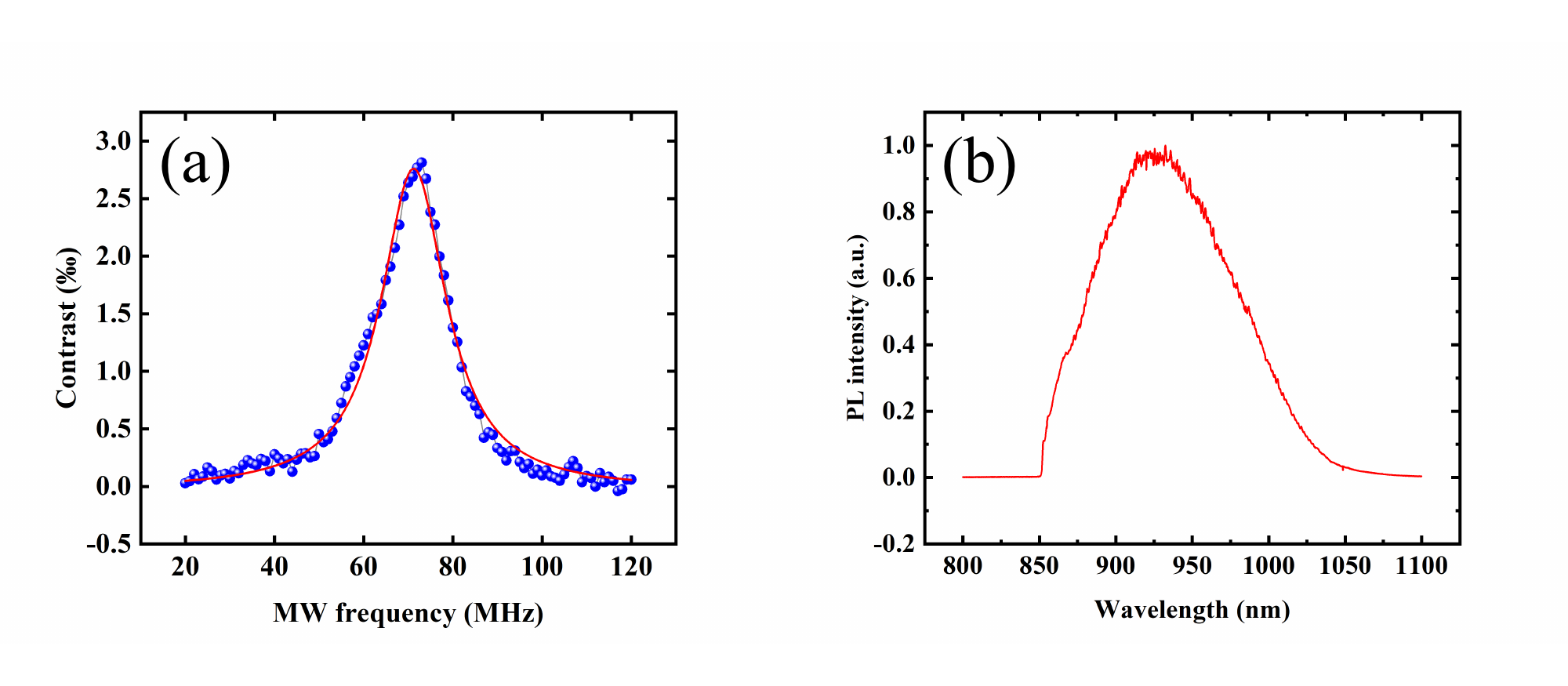}
		\caption{The characterization of generated V$_{Si}$ ensembles. (a) The ODMR measurement of V$_{Si}$ ensembles generated by an implantation dose of $1 \times 10^{13}$ $cm^{-2}$. The red line is the Lorentzian fitting of the data. The resonant frequency is $71.22 \pm 0.13$ MHz with a linewidth of $17.87 \pm 0.52$ MHz. (b) The corresponding room temperature PL spectrum of V$_{Si}$ ensembles.}
		\label{Figure 1}
	\end{figure*}
\section*{Experiment}

In our experiment, the sample was a commercially available high purity 4H-SiC epitaxy layer ($4^{o}$ off-axis, thickness is about 10 $\mu$m, with a nitrogen-doped concentration of $ 6 \times 10^{13} $cm$^{-3}$), which was diced to be 5 mm $\times$ 5 mm. We used a helium-ion microscope (ORION NanoFab, Zeiss) to generate V$_{Si}$ defects (see more details in section S1 in Supplementary information (SI)). The implantation energy was set to be 30 keV. By tunning the value of beam current to be 0.4 pA and varying the dwell time for implantation, we can precisely control the dose of FIB. There were two steps in the FIB implantation experiment. Firstly, we used the roaster scanning mode to generate a stripe-shaped ensemble as a mark with a high implantation dose of $1\times10^{13}$ ions cm$^{-2}$. Then we used the spot mode to generate defects arrays with various low implantation doses, from 100 ions/spot to 20 ions/spot (the lowest dose we can achieve for stable FIB implantation). There was no difference in the implantation process between the two working modes. A significant advantage of HIM is that the positions of generated defects are targeted and determined by a pre-designed file. Moreover, the spot size of the ion beam was sub-nanometer, with a diameter smaller than 0.5 nm. We inferred from the stopping and range of ions in matter (SRIM) simulation that the defects were located at about 179 nm beneath the surface of the epitaxy layer, with longitudinal and lateral straggling uncertainty of 47.4 nm and 59.3 nm, respectively (see section S2 in SI for more details). The implantation deviations were all below 60 nm in three dimensions. Moreover, the high-resolution HIM allows the ion beam to focus on the center of the two-dimensional patterns of nanostructures, and the implantation depth is near the center of thin-film structures with a hundred-nanometers thickness like nanophotonic waveguide\cite{Babin_2021} and nanobeams\cite{bracher2017selective}, which will be conducive to the integration of V$_{Si}$ defects into photonic structures with nanoscale precision and benefit the fluorescence extraction process. Previous works have investigated the relationship between ion implantation depth and the spin coherent time of NV centers in the diamond, showing a tenfold-enhanced spin coherent time with a nanometer-deeper implantation depth away from the surface noise\cite{PhysRevLett.113.027602,favaro2017tailoring,PhysRevX.9.031052}. Since the implantation depth is deeper than that of carbon ($\sim$ 60 nm) and silicon ($\sim$ 20 nm) ion implantation\cite{wang2017efficient,wang2017scalable,wang2019demand}, the spin coherent time is likely to be longer. To recover the lattice damage caused by implantation, the sample was annealed at 500 $ ^{o} $C in vacuum for 120 min with a heating ramp of 10 $^{o}$C $\cdot$ min$^{-1}$ and then freely cooled to room temperature. To remove the interstitial-related defects and also reduce the background fluorescence, we post-annealed the sample at 600 $ ^{o}$C in vacuum for 90 min.

\begin{figure*}[ht]
	\centering
	\includegraphics[scale = 0.4]{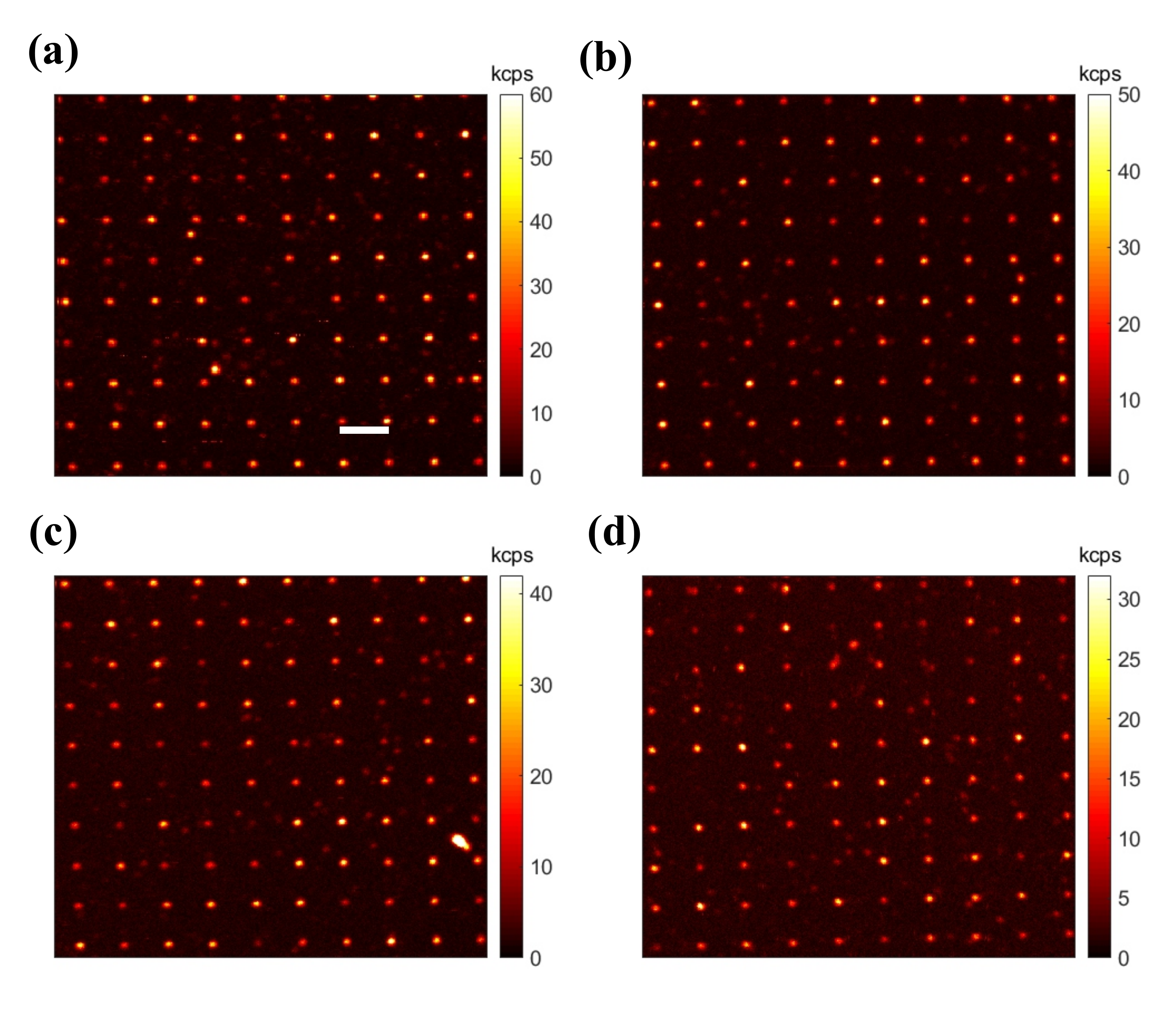}
	\caption{The confocal fluorescence scanning images at different FIB doses. The scale bar is 3 $\mu$m. (a) 100 ions/spot; (b) 80 ions/spot; (c) 60 ions/spot; (d) 40 ions/spot.}
	\label{fig:2}
	\label{fgr:example}
\end{figure*}

\section*{Results and Discussion}

We first characterized the V$_{Si}$ defects ensemble generated by high-dose FIB implantation. The sample was placed in a home-built confocal fluorescence scanning system and excited by a 730 nm laser through an infrared 0.85 N.A. objective (Olympus). To measure the room-temperature (RT) optically detected magnetic resonance (ODMR) spectrum, a 20-$\mu$m copper wire was paved on the surface of the epitaxy layer to deliver the radio-frequency microwave (MW), which was gated through a switch (ZASWA-2-50DR) and then amplified by a high-power amplifier (ZHL-20W-13+). The fluorescence was collected by a multimode fiber to a Femto (OE-200-Si). There are two nonequivalent silicon defects in 4H-SiC \cite{widmann2015coherent}, the hexagonal lattice site V$_{Si}$ defects (V$_{1}$) with good properties in low temperature\cite{nagy2018quantum,nagy2019high,morioka2020spin} and the cubic lattice site V$_{Si}$ defects (V$_{2}$). We focused on the V$_{2}$ defects for they can be manipulated at even room temperature \cite{widmann2015coherent,carter2015spin,simin2017locking}. Fig 1.(a) shows an ODMR contrast of about 0.28 $\%$ ,  with a linewidth (full width at half maximum) of $17.87 \pm 0.52$ MHz, which was similar to previous works \cite{wang2017efficient,wang2019demand,wang2017scalable}. The resonant frequency was $71.22 \pm 0.13$ MHz, corresponding to the zero-field-splitting $2D$ for V$_2$ silicon defects at room temperature \cite{widmann2015coherent}.  We further measured the RT photoluminescence (PL) spectrum of V$_2$ silicon defects using a spectrometer (iHR550, Horiba), as shown in Fig 1.(b). The observed spectrum was in the near-infrared range, which was similar to the previous results of the V$_{Si}$ ensemble and single V$_{Si}$ defect measured at room temperature \cite{widmann2015coherent,fuchs2015engineering,wang2017efficient,wang2019demand,wang2017scalable}.

\begin{figure*}[ht]
	\centering
	\includegraphics[scale = 0.5]{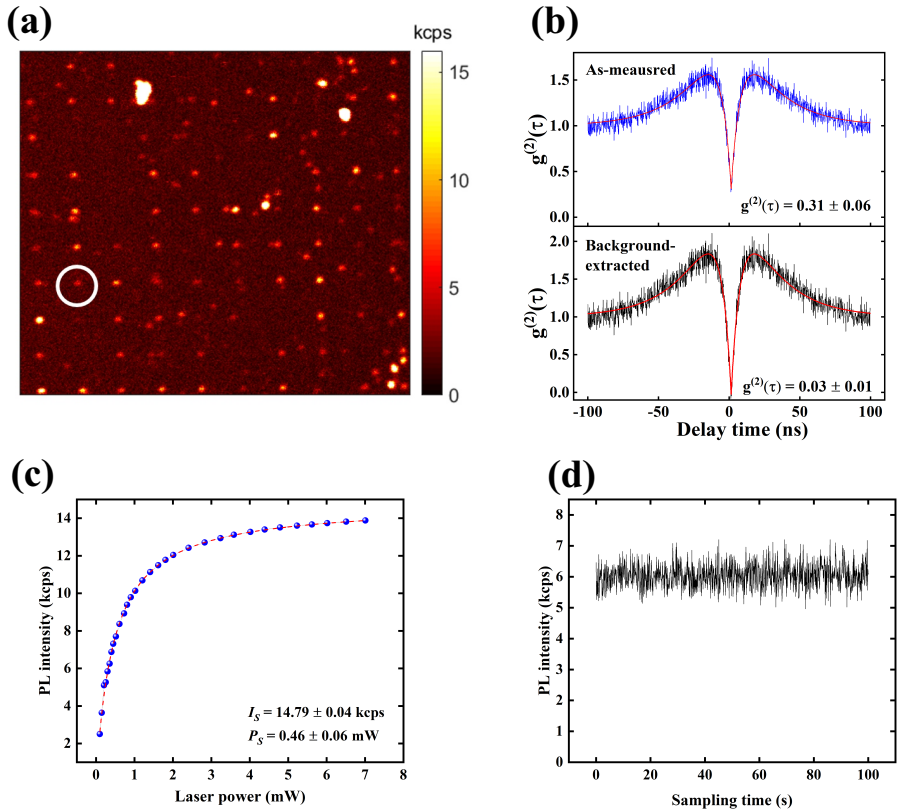}
	\caption{The characterization of generated single V$_{Si}$ defect. (a) The confocal fluorescence scanning image of generated V$_{Si}$ array in a dose of 20 ions/spot. (b) As measured (upper panel) and background-extracted (lower panel) second-order autocorrelation function of a single V$_{Si}$ defect under an excitation laser power of 0.5 mW. The red lines are the fitted data. (c) The photoluminescence intensity of a single V$_{Si}$ at different excitation laser powers, with a saturation laser power of $0.46 \pm 0.02$ mW and saturation PL intensity of $14.86 \pm 0.13$ kcps. (d) The time trace of photostability for a single V$_{Si}$ under an excitation laser power of 0.5 mW and a sampling time of 100 s.}
	\label{fig:3}
	\label{fgr:example}
\end{figure*}

We proceeded to perform the experiment on the V$_{Si}$ defect arrays generated by the FIB with low implantation doses. We optimized the dwell time and beam current of the HIM to change the implantation dose per spot. We found that the dwell time has a linear relation with the dose at a certain current (see table S1 in SI for more details). The fluorescence of V$_{Si}$ defects were collected by a 1.3 N.A. oil objective (Nikon) to enhance the collection efficiency, then separated by a 50:50 beam splitter and coupled to two avalanche photodiodes (APDs) with a 40-$\mu$m-diameter pinhole. The confocal scanning images under an excitation laser power of 0.5 mW for defect arrays generated at doses ranging from 100 to 40 ions/spot are shown in Fig 2. For each 10 $\times$ 10 array, the spots were separated by 3 $\mu$m. The fluorescence intensity for each spot decreased significantly with the dose reduction. The corresponding statistics of generated V$_{Si}$ defects per spot at different doses are shown in section S3 in SI. As described below, the generated rate of a single V$_{Si}$ reduced exponentially with the dose. We then further reduced the dose to increase the generation rate of single V$_{Si}$ defects.

When the FIB dose reduced to 20 ions/spot (the lowest dose that HIM  can achieve), the array was still obvious, as shown in Fig 3.(a). However, the fluorescence intensity for a particular spot was comparably weak, which was similar to the reported intensity of a single V$_{Si}$ \cite{wang2017efficient, wang2019demand, wang2017scalable}. To identify single V$_{Si}$ defect centers, we measured the second-order autocorrelation functions ($g^{2}(\tau)$) of the generated color centers ($\tau$ is the time delay between two paths). We present the result of the single V$_{Si}$ defect in the white circle in Fig 3.(a) as an example. When excited by a laser power of 0.5 mW, the signal counts $S = 6$ kcps (kilo counts per second) and the background counts $B = 2$ kcps. Since the intensity of background was comparable with signal, the as-measured data normalized data ($C_N(\tau)$) was corrected using $g^2(\tau) = \frac{(C_N(\tau)-(1-\rho^2))}{\rho^2}$, where $\rho = \frac{S}{S+B}$. As it is shown in Fig 3.(b), we fitted both the as-measured (upper panel) and background-extracted (lower panel) data with $g^2(\tau) = 1 - (1+a) \exp(-\frac{|\tau|}{\tau_1}) + b\exp(-\frac{|\tau|}{\tau_2})$, where $a$, $b$, $\tau_1$ and $\tau_2$ are fitting parameters. The fitted $g^2(0)$ were $0.31 \pm 0.01$ and $0.03 \pm 0.01$, respectively, both were far below 0.5, verifying that it is a single V$_{si}$ defect. We then measured the PL intensity $I$ as a function of excitation laser power $P$ (Fig 3.(c)). The red line is the fit using $I(P)=\frac{I_S}{I+\frac{P_S}{P}}$, where $I_S$ is the maximum intensity, and $P_S$ is the saturation laser power. Inferred from the fit, $P_S = 0.47 \pm 0.02$ mW and $I_S = 14.86 \pm 0.13$ kcps, respectively. Since the stability of a single-photon source is also crucial in quantum information applications, we traced the PL intensity of the V$_{Si}$ defect under an excitation laser power of 0.5 mW with a sampling time of 100 s, as shown in Fig 3.(d). The intensity kept stable with no blinking or bleaching, showing that the synthesized single V$_{Si}$ defect has excellent photostability.

\begin{figure*}[ht]
	\centering
	\includegraphics[scale = 0.5]{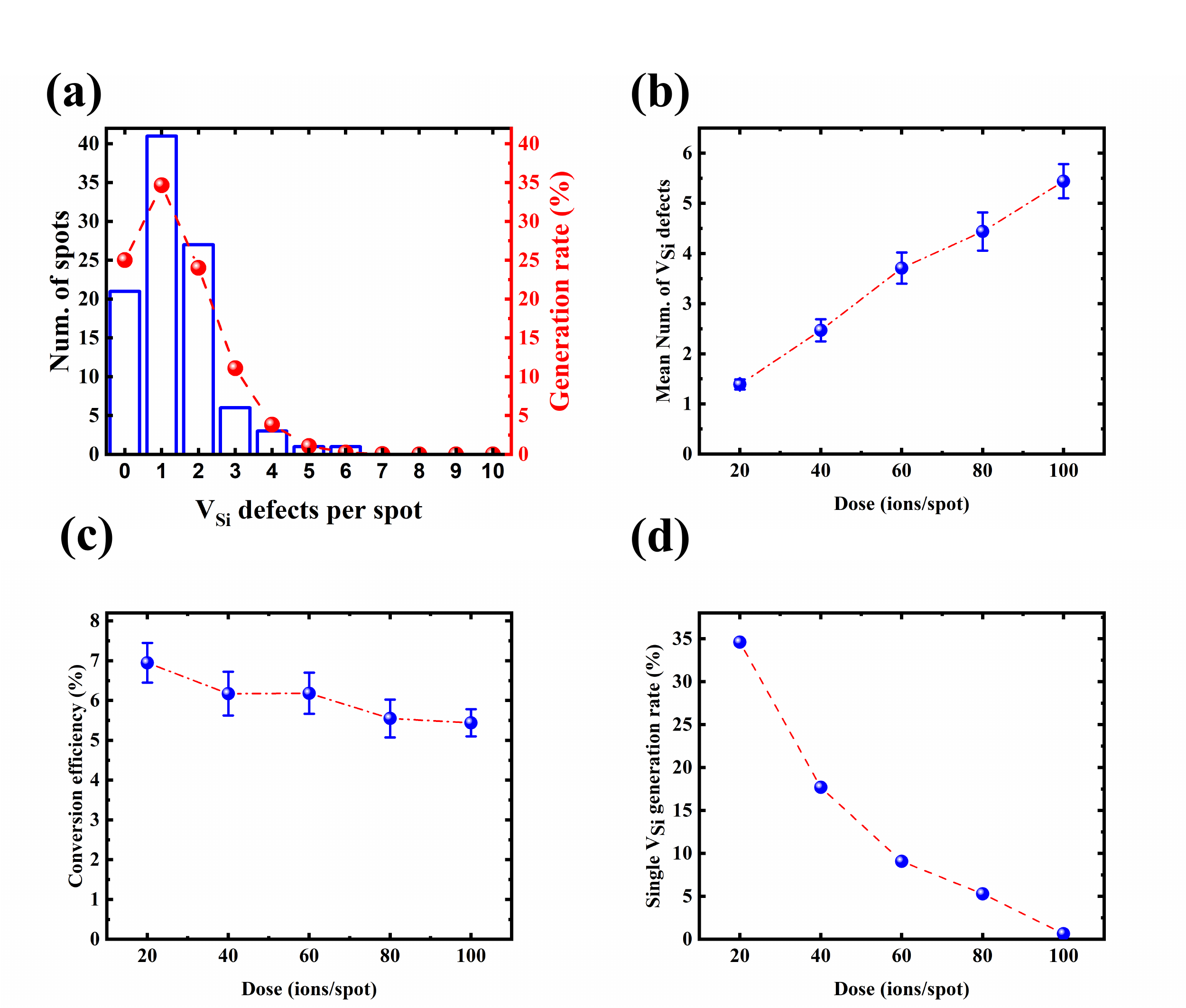}
	\caption{The statistics of generated  V$_{Si}$ at different implantation doses. (a) The statistics of generated V$_{Si}$ at a dose of 20 He$^+$ ions/spot. (b) The average number of generated V$_{Si}$ defects per spot as a function of dose (ions/spot). (c) The conversion yields of He$^+$ to V$_{Si}$ defects as a function of dose (ions/spot). (d) The generation rate of a single V$_{Si}$ defect as a function of dose (ions/spot).}
	\label{fig:4}
	\label{fgr:example}
\end{figure*}

To analyze the effect of implantation dose, we systematically counted and estimated the number of generated V$_{Si}$ defects in each spot at different doses with a sampling size of 100 spots. The statistic method used was similar to the previous works\cite{Babin_2021,chen2019laser,chen2017laser}. By combining fluorescence intensity detection and measuring $g^2(0)$, we found that the number of V$_{Si}$ defects per spot could be estimated by a fluorescence intensity unit of 8 kcps, which was approximately the intensity of one V$_{Si}$ defect under an excitation laser power of 0.5 mW (see section S3 in SI for the methodology to estimate the number of color centers per spot). The mean number of V$_{Si}$ defects and the inferred generation rates were fitted by a Poisson distribution function $P(\lambda) = \frac{e^{-\lambda}\lambda^k}{k!}$, where $\lambda$ is the mean number of V$_{Si}$ defects per spot. As a representative, the statistics of V$_{Si}$ defects generated at a dose of 20 ions/spot are shown in Fig 4.(a). The blue histograms are the counting results, and the red line is the Poisson fit. At a dose of 100 ions/spot, up to 10 V$_{Si}$ defects can be generated within a spot, with $\lambda = 5.442 \pm 0.340$. It was hard to find single V$_{Si}$ defects (the number of single V$_{Si}$ defects within 100 spots is usually smaller than 3) until the FIB dose was reduced to 40 ions/spot, with 17 single V$_{Si}$ defects to be identified in the 10 $\times$ 10 arrays (Fig 2.(a)). In the lowest FIB dose of 20 ions/spot, 41 single V$_{Si}$ defects occurred in the 10 $\times$ 10 array, with $\lambda = 1.39 \pm 0.10$. The inferred generation rate for a single V$_{Si}$ defect was $\sim$ 35 $\%$. With an average implantation dose of 20 ions/spot, the conversion yield of He$^+$ ions to V$_{Si}$ defects was estimated to be 6.95 $\%$, which was similar to the former He$^+$ ions implantation experiment \cite{Babin_2021}. Similar analysis can be applied to cases of other doses. The mean number of V$_{Si}$ defects per spot and the conversion yields as a function of dose are visible in Fig 4.(b) and (c), respectively. The conversion yield increased with the reduction of implantation dose, from 5.44 $\%$ to 6.95 $\%$. Such a phenomenon has been observed in the Si$^{2+}$ FIB experiment in SiC \cite{wang2017scalable}. Also, the inferred single V$_{Si}$ defect generation rate increased as the dose decreased, as shown in Fig 4.(d). The generation rate of a single V$_{Si}$ defect ($\sim$ 35 $\%$) was comparable with previous works of mask-assisted helium-ion, carbon-ion implantation and Si$^{2+}$ FIB experiment\cite{Babin_2021,wang2017efficient,wang2019demand,wang2017scalable}. However, a drawback of carbon and silicon ion implantation is that the residual lattice damage is significant, which degrades the spin properties of spin defects. Such a phenomenon has been observed in nitrogen-ion-implanted NV centers in the diamond \cite{favaro2017tailoring}. Compared with carbon and silicon ions, helium has a smaller atomic mass, which causes smaller damage and a deeper implantation depth when accelerated by the same kinetic energy. The charge state of induced spin defects is inferred to be more stable due to the gentle damage caused by helium ions. Furthermore, the effects of surface noise will degrade with a deeper implantation depth and prospect and longer spin coherent time, inferred from the case of NV centers in the diamond\cite{PhysRevLett.113.027602,favaro2017tailoring,PhysRevX.9.031052}.

\section*{Conclusion}
In conclusion, we investigated a maskless and targeted method for precisely generating single V$_{Si}$ defects arrays in 4H-SiC using a high-resolution helium-ion microscope. ODMR and PL spectra measurements confirmed the types of generated defects. The fluorescence properties of single V$_{Si}$ defects were characterized, showing excellent single-photon properties, photostability, and saturation intensity of $\sim$ 15 kcps. The effects of FIB dose were systematically studied, and the conversion yield and generation rate of single V$_{Si}$ defects rise with the reduction of the dose. A conversion yield of $\sim$ 6.95 $\%$ and a generation rate of $\sim$ 35 $\%$ for a single V$_{Si}$ defect were realized at a dose of 20 ions/spot. Due to the high generation rate of the single defect and the precise control of implantation position, this work is meaningful for the applications in integrating color centers to semiconductor-based microstructures like waveguides, nanopillars, and photonic crystal cavities \cite{Babin_2021,radulaski2017scalable,bracher2017selective}, precise quantum sensing based on optically-addressable spin defects \cite{kraus2014magnetic,simin2015high,niethammer2016vector}, and thus benefits the quantum information process on spin-to-photon interface \cite{morioka2020spin, Babin_2021, soykal2016silicon}.
\section*{Supplementary Information}
\subsection*{S1. The helium-ion microscope}

The helium-ion microscope (HIM) components are shown in Fig 5. There is a single crystal tungsten trimer composed of three tungsten atoms in gas helium with a low pressure of 10$^{-6}$ Torr. Helium atoms are ionized into three beam currents by the high voltage of the trimer. One of the beam currents will be selected and extracted by the extractor and focused on the ion beam. After passing an aperture of 20 $\mu$m, it will be deflected and focused into a spot with a sub-0.5 nm diameter, scanning the target surface. The E-T detector captures the secondary electrons excited by the incident ions spot by spot to realize a 0.5 nm resolution image, which is helpful for the on-demand FIB implantation and the integration of color centers to nanostructure.	
\begin{figure}[htbp]
	\centering
	\includegraphics[scale = 0.4]{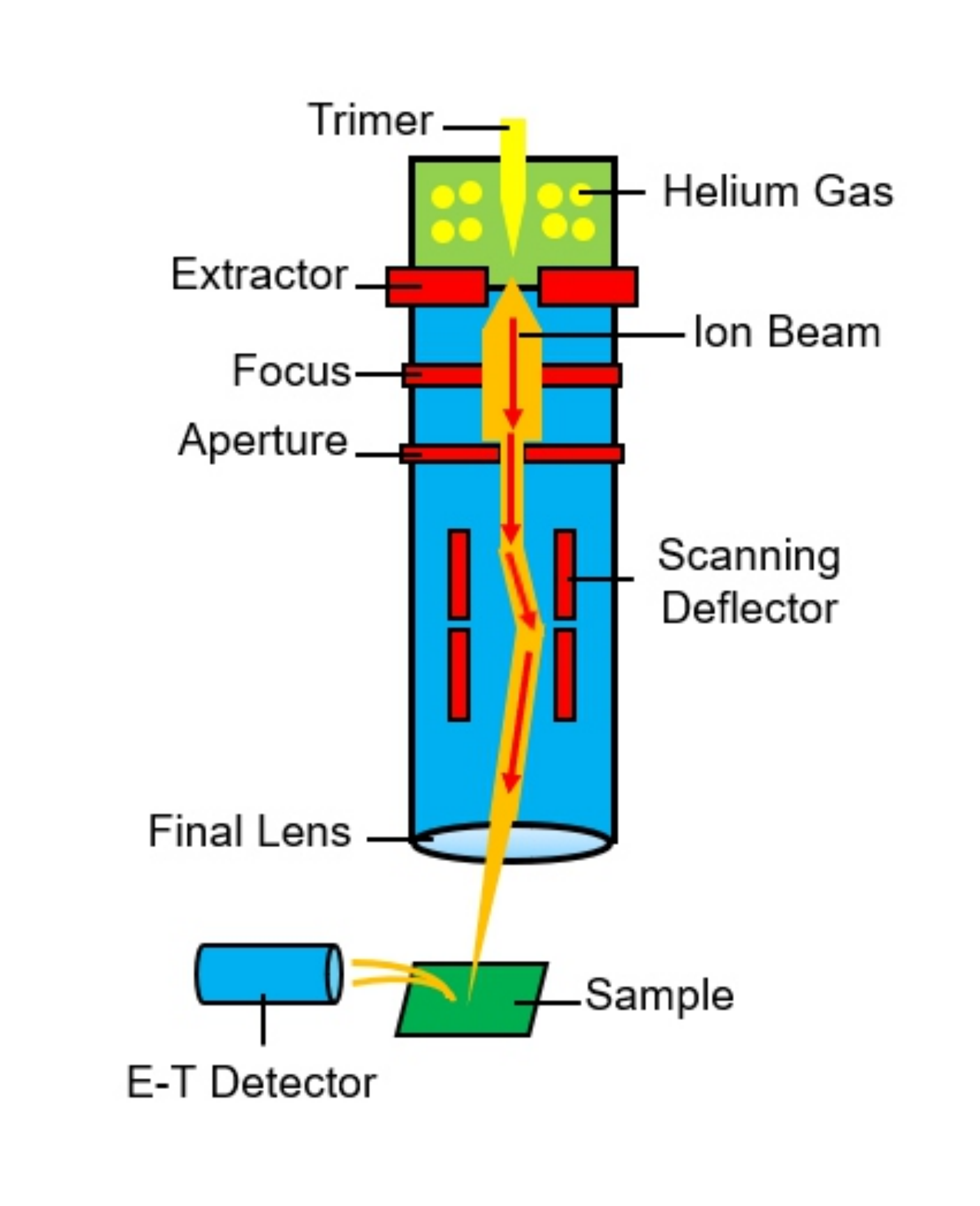}
	\caption{The structure of the helium-ion microscope.}
	\label{Figure S1}
\end{figure}

There are two kinds of implantation modes of HIM. The first one is the spot mode. The helium ion beam is focused within a diameter smaller than 0.5 nm in this mode. The implantation doses are controlled by the beam current and dwell time on each spot. The second mode is roaster scanning mode. Implantation can be realized within a specific geometry after setting the number of ions in the unit area (ions/cm$^2$), slightly different from spot mode with a dose unit of ions/spot. We use scanning mode to generate V$_{Si}$ defects ensemble strip and spot mode to generate V$_{Si}$ defects arrays.

It is noticed that the perturbation of beam current from the set value (0.4 pA) is within the tolerance when HIM is working. As shown in TABLE S1, the dwell time has a linear relation with the dose with a particular beam current. The accuracy of incident ions is determined by the accuracy of HIM’s dwell time, which is about 0.1 $\mu$s.  With a beam current of about 0.4 pA, a dwell time of 0.1 $\mu$s determines a dose deviation of about 0.25 ions/spot. 	
\begin{table}
	\caption{The relation between dose, beam current and dwell time}
	\label{tbl:example}
	\begin{tabular}{lll}
		\hline
		Dose (ions/spot)  & Dwell time ($\mu$s) & Beam current (pA)   \\
		\hline
		100   & 37.6 & 0.420   \\
		\hline
		80 & 31.2 & 0.419  \\
		\hline
		60  & 23.8 & 0.423  \\
		\hline
		40 & 15.6 & 0.417 \\
		\hline
		20 & 7.9 & 0.412 \\
		\hline
	\end{tabular}
	\label{Table S1}
\end{table}

\subsection*{S2. SRIM simulation of the depth distribution of He$^+$-ions-generated silicon vacancies}
\begin{figure}[htbp]
	\centering
	\includegraphics[scale = 0.4]{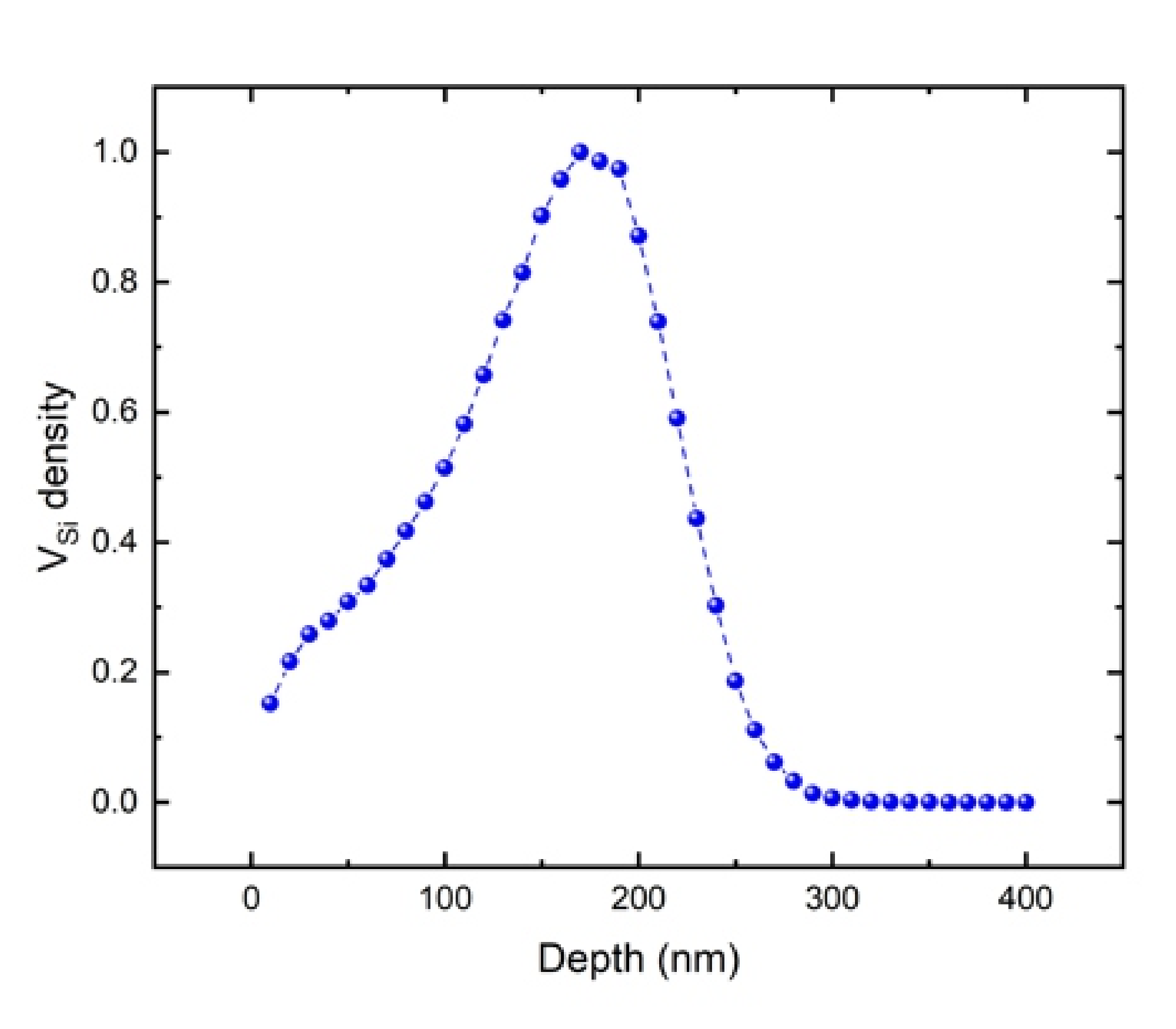}
	\caption{SRIM simulated depth distribution of 30 keV-He$^+$-generated silicon vacancies.}
	\label{Figure S2}
\end{figure}
We use the stopping and range of ions in matter (SRIM) simulation to simulate the process of 30 keV He$^+$ ions implantation. As shown in Fig 6, the simulated implantation depth is about 179 nm, with a longitudinal straggling of 47.4 nm.	

\begin{figure}[htbp]
	\centering
	\includegraphics[scale = 0.35]{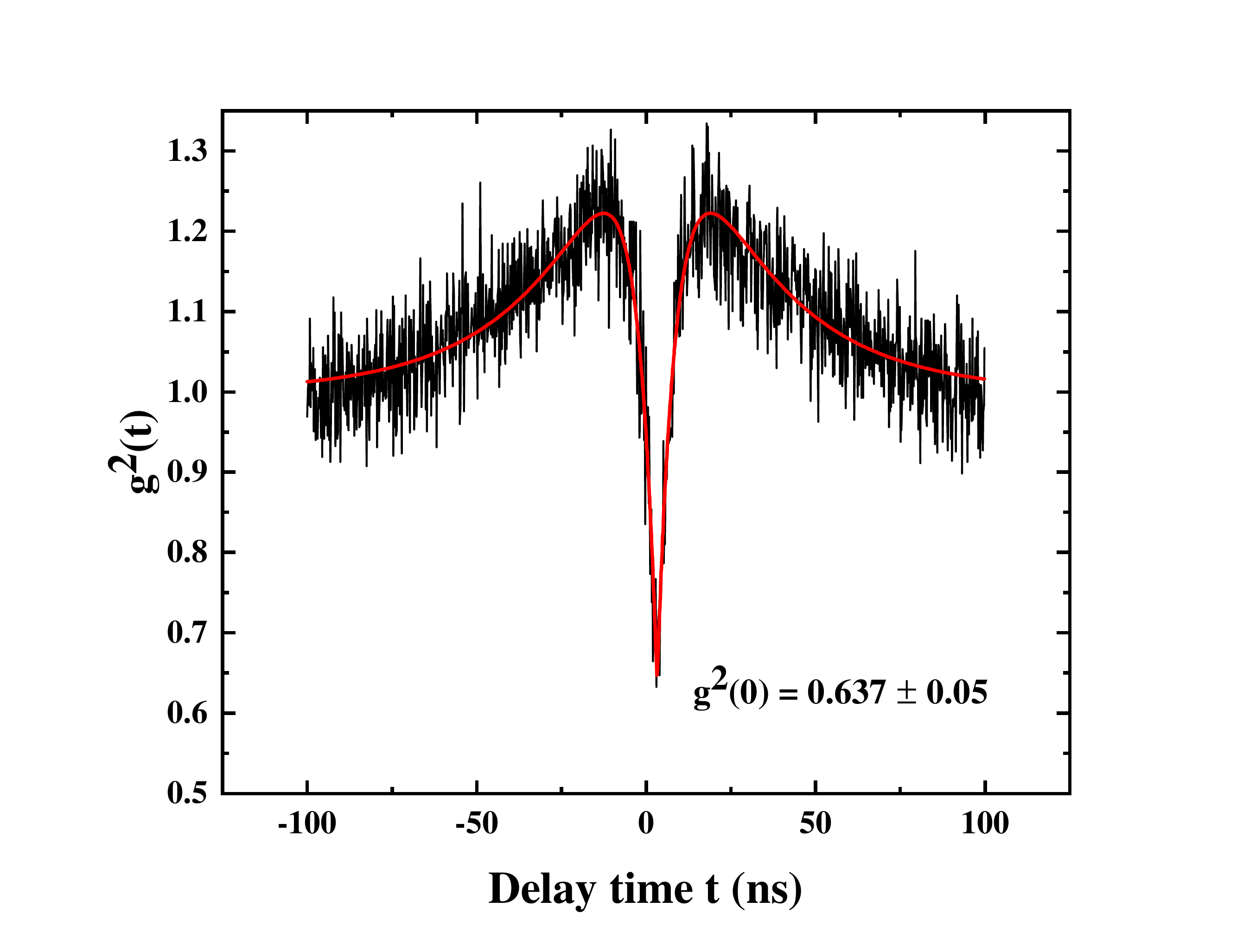}
	\caption{The second-order autocorrelation function of a typical double-photon source spot with a luminescence intensity of 13.5 kcps.}
	\label{Figure S3}
\end{figure}

\subsection*{S3. The analysis of V$_{Si}$ defects generation rate}
\begin{table*}
	\caption{The relation between luminescence intensity, $g^2(0)$ and number of V$_{Si}$ centers per spot}
	\label{tbl:example}
	\begin{tabular}{lll}
		\hline
		Luminescence intensity under 0.5 mW (kcps)  & $g^2(0)$ & Number of V$_{Si}$ defects  \\
		\hline
		4 $\sim$ 8   & 0 $\sim$ 0.32 & 1   \\
		\hline
		8 $\sim$ 16 & 0.32 $\sim$ 0.65 & 2  \\
		\hline
	\end{tabular}
	\label{Table S2}
\end{table*}
We estimate the number of V$_{Si}$ centers by combining the HBT experiment and measuring the luminescence intensity. We measured the second-order autocorrelation functions $g^2(0)$ of picked color centers within a 10 $\times$ 10 array under a laser power of 0.5 mW, which were clarified into different groups by luminescence intensity levels. And every group contains at least four color centers. We found that the luminescence intensity and $g^2(0)$ satisfied the relation in the table below. 

According to the previous work, color centers with a $g^2(0)$ from 0 to 0.32 are identified as single-photon sources, and color centers with $g^2(0)$ from 0.33 to 0.65 are clarified as double-photon sources. We measured several spots with intensity from 4 to 8 kcps and from 8 to 16 kcps, respectively, all the measured color centers satisfied the relationship in Table S2. In Fig 7, we showed the second-order autocorrelation function of a spot with a luminescence intensity of 13.5 kcps as a typical double-photon source, with $g^2(0) = 0.637 \pm 0.05$. Hence, we can assume that an intensity unit of 8 kcps corresponds to one silicon vacancy. Since the luminescence intensity of a single silicon vacancy is low, it will be time-consuming to measure $g^2(0)$ of all the spots, especially those containing 3 silicon vacancies or more, whose number of color centers can be estimated by the assumed intensity unit. For example, a spot with a luminescence intensity of 23 kcps contains 3 V$_{Si}$ defects and a spot with a luminescence intensity of 46 kcps contains 6 V$_{Si}$ defects.
\begin{figure}[htbp]
	\centering
	\includegraphics[scale = 0.3]{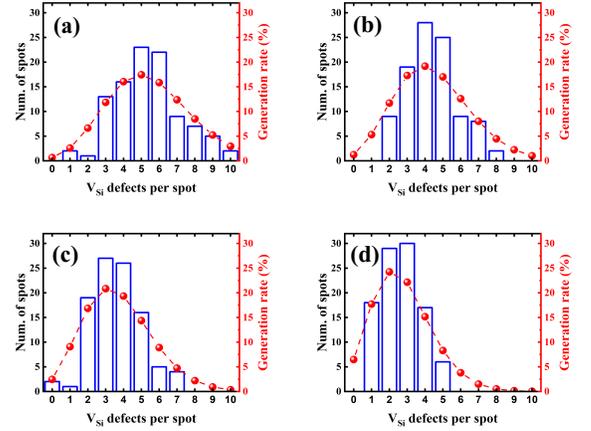}
	\caption{The statistics of generated V$_{Si}$ defects per spot at different doses. (a) 100 ions/spot. (b) 80 ions/spot. (c) 60 ions/spot. (d) 40 ions/spot.}
	\label{Figure S4}
\end{figure}
At each dose, we statistic the number of spots with a corresponding number $N$ of generated V$_{Si}$ defects per spot in a 10 $\times$ 10 array, shown in blue histograms in Fig 8. By normalizing the number $N$ and using the Poisson fitting function $P(\lambda) = \frac{e^{-\lambda}\lambda^k}{k!}$ mentioned in the main text, we can infer the possibility to generate $N$ ($N = 0 \sim 10$) V$_{Si}$ defects per spot, which is equal to the generation rate (Red lines and dots in Fig 8). The statistics of generated V$_{Si}$ defects from 100 He$^+$ ions/spot to He$^+$ 40 ions/spot are shown in Fig 8, and the statistics of generated V$_{Si}$ defects at 20 He$^+$ ions/spot are shown in Fig 4.(a) in the main text.

\section*{Conflicts of Interest}
	There are no conflicts to declare.
	
\section*{Acknowledgments}
	
	The FIB experiment was carried out at the Center for Micro and Nanoscale Research and Fabrication at University of Science and Technology of China. This work was supported by the Innovation Program for Quantum Science and Technology (Grants No. 2021ZD0301400), the National Natural Science Foundation of China (Grants No. 61725504, No. 11774335, No. 11821404, and No. U19A2075), the Anhui Initiative in Quantum Information Technologies (Grants No. AHY020100, and No. AHY060300), the Fundamental Research Funds for the Central Universities (Grant No. WK2030380017), and the China Postdoctoral Science Foundation (Grants No. BX20200326 and 2021M693099).


\end{document}